\documentstyle[aps]{revtex}

\begin{document}
\preprint{Applied Physics Report 94-39}
\title{Mesoscopic Conductance Oscillations in a Normal Sample Controlled by the
Superconductivity of an NS Boundary}
\author{ A. Zagoskin\thanks{Email: alexz@fy.chalmers.se}, A.
Kadigrobov\thanks{permanent address: B.I. Verkin Institute for Low Temperature
Physics and Engineering, 47 Lenin Ave., 310 164 Kharkov, Ukraine},  R.I.
Shekhter\thanks{Email: shekhter@fy.chalmers.se}, and M. Jonson\thanks{Email:
jonson@fy.chalmers.se}}
\address{Department of Applied Physics, Chalmers University of Technology and
G\"{o}teborg University, S-412 96 G\"{o}teborg, Sweden}
\maketitle
\draft

\begin{abstract}
An arrangement of mesoscopic experiments is considered when the conductance of
a mesoscopic normal metal sample is governed not only by the magnetic flux
threading the sample but by the supercurrent inside the superconductor coupled
to the normal metal sample as well. The conductance of the system is shown to
be associated with the SNS Josephson current $j_J$ through the normal part of
the system.\\
By way of example we consider a model when quasi1D wire is connected to a
superconductor by its ends and coupled to two reservoirs of electrons by the
normal metallic leads. The conductance oscillations caused by change of
magnetic field and the supercurrent inside the superconductor have been
calculated.\\
The possibility  of direct observation of the Josephson current through
measurement of the thermal noise in the normal conductor is suggested.
\end{abstract}


\section{Introduction}

Quantum electronic transport in mesoscopic systems has been a subject of many
experimental and theoretical investigations, the conductance oscillations and
fluctuations being usually governed by the magnetic flux threading the sample
(see for review Ref.\cite{1} and references therein). Recently much attention
was attracted by the mesoscopic systems containing normal-superconducting
interfaces, due to nontrivial properties of normal transport in these systems
\cite{Beenakker,Lambert,Petrashov,trombone,2}.

Here we consider a situation where the normal conductance oscillations  are
controlled not only by the magnetic flux through the normal metal mesoscopic
sample,  but by the supercurrent inside the superconductor coupled to it. In
this case the oscillations are associated with the SNS Josephson current
passing through the normal part of the system.

We discuss here the following model of a normal-superconducting mesoscopic
system (Fig.\ref{f.1}). The normal leads are modelled by ideal normal wires.
The scattering processes are confined to NS boundaries and T-junctions. The
latter are described by real S-matrices. The system is coupled to a
superconductor, S, and to two equilibrium electronic reservoirs, L and R.

The main result of this paper is as follows. In the ballistic  limit (no
impurity scattering and no mode mixing) and in the absence of the normal
reflection at NS-boundaries the normal conductance of the system  is given by
the formula
\begin{equation}
G = \frac{2e^2}{h} N_{\perp} \frac{ 2\epsilon^2}{\cos \phi_J +
 1 + 2\epsilon^2} + o(N_{\perp}). \label{main1}
\end{equation}
Here $\epsilon$ is proportional to the probability of a particle leaving the
normal part at a junction connecting it to a reservoir, $N_{\perp}$ is the
number of transverse modes in the lead ADCB (of length $\cal L$ and crossection
$\sim N_{\perp}\lambda_F^2$), and $\phi_J$ is the gauge invariant phase
difference between the points A and B of the superconductor.
In the limit $\epsilon \rightarrow 0$ this equation can
be rewritten as follows (cf. \cite{S})
\begin{equation}
   G =
\epsilon \left[ N_{\perp}\left.\frac{e^2}{\pi  \hbar }\right.  -
\left.\frac{e {\cal L} }{
 \hbar  \bar{v}_{F}}\right. \frac{d j_J^{(\epsilon)}}{d \phi_J} \right]+
o(N_{\perp}), \label{main}
\end{equation}
where  $\bar{v}_F = N_{\perp}^{-1}
\sum_{\nu=1}^{N_{\perp}}v_{F,\nu}^{\parallel}$ is the average longitudinal
Fermi velocity, $j_J^{(\epsilon)}$ is the Josephson current through
ADCB [weakly connected  to the normal reservoirs], controlled by the phase
difference $\phi_J$.

This relation allows us to measure directly the Josephson current vs. phase
dependence simply by measuring the equilibrium current fluctuations in a
system.

 \section{Basic equations}

In our analysis, we assume that the size of normal part of the system, $\cal
L$, is less than both the phase breaking length, $L_{\phi} =
(D\tau_{\phi})^{1/2}$, and the normal metal coherence length, $L_T = ( \hbar
D/k_B T)^{1/2}$ (but exceeds the superconducting coherence length $\xi_0$).
Here $\tau_{\phi}^{-1}$ is the inelastic scattering rate, $D$ is the diffusion
constant of quasiparticles, $k_B$ is the Boltzmann constant, and $T$ is
temperature.  For the time being, we put the number of normal modes,
$N_{\perp}$, to unity. As usually, when discussing a coherent transport at a NS
boundary, we use the two-component wave function \cite{Beenakker} of the
quasiparticle in normal part of the system:
\begin{equation}
{\bf v} = \left( \begin{array}{l} v^e \\ v^h \end{array} \right), \label{2}
\end{equation}
where $v^{(e,h)}$ describes an electron-(hole-)like quasiparticle (with a given
energy $E$). Due to one-dimensionality of the system, the coordinate dependence
of these amplitudes is
\begin{equation}
v^e(x) = v^e(0) e^{ipx};\:\:\: v^h(x) = v^h(0) e^{-iqx},
\end{equation}
where
$p, q$ are momenta of electron and hole respectively.

In the system we are discussing the scattering occurs only at the junctions (C,
D) and NS boundaries (A, B). The latter includes both  normal and Andreev
scattering, which are characterized by the corresponding reflection amplitudes,
$\rho_{e,h}$ and $\beta_{\pm}$. The quantity $\rho_{e,h}$ is the probability
amplitude for an electron-(hole-)like excitation incident from the normal lead
to be reflected back as an electron (hole) (normal reflection), while
$\beta_{\pm}$ is the probability amplitude of its conversion to the hole
(electron) (Andreev reflection). The latter  amplitudes depend explicitly on
the superconducting phase in the reflection point.

The scattering matrices in the junctions relate the amplitudes of
quasiparticles
in the three connected leads (see Fig.\ref{f.2}a):
\begin{equation}
{\bf w}_{\alpha} = \sum_{\beta=1}^3{\cal S}_{\alpha\beta} {\bf v}_{\beta};
\:\:\alpha,\beta = 1,2,3.
\end{equation}
Here ${\bf v}_{\alpha}$ (${\bf w}_{\alpha}$) is the in(out)going wave in the
$\alpha$-th lead respectively, described by a two-component wector
(Eq.(\ref{2})); ${\cal S}_{\alpha\beta}$ is a $3 \times 3$ unitary matrix, each
element of it being a $2 \times 2$  diagonal matrix (since no electron-hole
mixing occurs in the normal part of the system):
\begin{equation}
{\cal S}_{\alpha\beta} = \left( \begin{array}{ll} {\cal S}_{\alpha\beta}^e &
0\\ 0 & {\cal S}_{\alpha\beta}^h \end{array} \right).
\end{equation}

Now we can easily exclude the "stubs" AC and BD from   consideration, using the
boundary conditions at the NS boundary, which in matrix form read (see
Fig.\ref{f.2}a)
 \begin{eqnarray}
{\bf v}_3 = {\bf A}\:{\bf w}_3; \:\:\:
{\bf A} = \left( \begin{array}{ll} \rho_- & \gamma_-\\
 & \\  \gamma_+ & \rho_+  \end{array} \right).
\end{eqnarray}
 Here the dot denotes the matrix multiplication   in $(e,h)$-space,  $\rho_- =
\rho_e \exp(2ipl)$, $\rho_+ = \rho_h \exp(-2iql)$,
$\gamma_- = \beta^*_- \exp(i(p-q)l)$,  $\gamma_+ = \beta_+ \exp(i(p-q)l)$, and
$l$ is the length of the "stub" (AC or BD).

As a result, instead of initial $3 \times 3$ "supermatrix", $\cal S$, we obtain
a $2 \times 2$ one, $\bf S$, which relates only the wave amplitudes in the lead
directly connecting the normal reservoirs. The matrix elements of the two are
related as follows:
\begin{equation}
{\bf S}_{ab} = {\cal S}_{ab} + {\cal S}_{a3} \cdot{\bf B}\cdot {\cal S}_{3b};
\:\:\: a,b = 1,2,
\end{equation}
 where the 2$\times$2-matrix in the $(e,h)$-space, $\cal B$, is given by
\begin{eqnarray}
{\bf B} = \left\{ ({\cal S}_{33}^e - \frac{\rho_+}{\Gamma})(S_{33}^h -
\frac{\rho_-}{\Gamma}) -
\frac{\gamma_+\gamma_-}{\Gamma^2}\right\}^{-1} \left( \begin{array}{ll}
S_{33}^{h} - \frac{\rho_-}{\Gamma} & -\frac{\gamma_-}{\Gamma} \\
 & \\ -\frac{\gamma_+}{\Gamma} & S_{33}^{e} - \frac{\rho_+}{\Gamma}
\end{array}\right);
\end{eqnarray}
Here \begin{eqnarray}
\Gamma = \gamma_+\gamma_- - \rho_+\rho_-.
\end{eqnarray}

We will regard the case
of weak coupling of the system to the normal reservoirs. Moreover,
we  choose  the initial scattering matrices, $\cal S$,  to be real and
one-parametric \cite{Buttiker}, and satisfying electron-hole symmetry, so that
${\cal S}^e = {\cal S}^h$.  Then the "left" and "right" initial scattering
matrix,  in the junctions C and D  respectively, have the form :
\begin{eqnarray}
{\cal S}_L = \left(\begin{array}{lll}
-\sqrt{1-2\epsilon} {\bf I} & \sqrt{\epsilon} {\bf I} & \sqrt{\epsilon} {\bf I}
\\
& & \\
\sqrt{\epsilon} {\bf I} & r {\bf I}& t {\bf I}\\
& & \\
\sqrt{\epsilon} {\bf I} & t {\bf I} & r {\bf I}
\end{array} \right);\:\:\:
{\cal S}_R = \left(\begin{array}{lll}
r {\bf I} & \sqrt{\epsilon} {\bf I} & t {\bf I}\\
& & \\
\sqrt{\epsilon} {\bf I} & -\sqrt{1-2\epsilon} {\bf I} & \sqrt{\epsilon} {\bf
I}\\
& & \\
t {\bf I} & \sqrt{\epsilon} {\bf I} & r {\bf I}
  \end{array} \right).
\end{eqnarray}
Here $0 \leq \epsilon \leq 1/2$ is a  coupling parameter, \\
$r = (1/2)\left[(1-2\epsilon)^{1/2} -1\right], \quad
t = (1/2)\left[(1-2\epsilon)^{1/2} +1\right]$ are the reflection (transmission)
amplitudes between the "stub" and the lead CD,\\
{\bf I} is the unit $2\times 2$-matrix in $(e,h)$-space.

In the absence of normal scattering at the NS boundary ($\rho_{e,h} = 0$)
this  leads to the following expressions for the effective scattering matrices:
\begin{eqnarray}
{\bf S}_L = \left(\begin{array}{ll} {\bf Q}_L & {\bf T}_L\\
{\bf T}_L & {\bf R}_L\end{array}\right); \:\:\:
{\bf S}_R = \left(\begin{array}{ll} {\bf R}_R & {\bf T}_R\\
{\bf T}_R & {\bf Q}_R\end{array}\right); \label{EFMAT}
\end{eqnarray}
Here the submatrices have the following form:
\begin{eqnarray}
{\bf T} = \sqrt{\epsilon}{\bf I} + \frac{\sqrt{\epsilon}t}{r^2 -
\gamma_+^*\gamma_-^*} \left( \begin{array}{ll} r & -\gamma_+^* \\
-\gamma_-^* & r \end{array}\right); \\
{\bf Q} = -\sqrt{1-2\epsilon}{\bf I} + \frac{\epsilon}{r^2 -
\gamma_+^*\gamma_-^*} \left( \begin{array}{ll} r & -\gamma_+^* \\
-\gamma_-^* & r \end{array}\right); \\
{\bf R} =  r{\bf I} + \frac{t^2}{r^2 - \gamma_+^*\gamma_-^*} \left(
\begin{array}{ll} r & -\gamma_+^* \\
-\gamma_-^* & r \end{array}\right).
\end{eqnarray}
Notice that the effective scattering matrices do mix electron- and holelike
parts of the quasiparticle wave function, since they include the effects of
Andreev scattering at NS boundaries.

\section{Normal conductance of the system }

After  substituting the initial scattering matrices by the effective ones
(\ref{EFMAT}), we  reduce the initial problem to   a generalized Landauer
one\cite{Imry}, where a 1D normal lead with scatterers connecting two
equilibrium electronic reservoirs (Fig.\ref{f.2}c). The difference from the
generic Landauer case is due to the  necessity to account for electrons and
holes separately   \cite{Lambert}.
 The normal  two probe conductance between the  reservoirs at   temperature $T
\rightarrow 0$ is thus given by the Lambert's formula \cite{Lambert,Lambert2}
\begin{eqnarray}
 G = \frac{2e^2}{h} \left\{ \left< T^>_0 \right> + \left< T^>_a \right> +
\frac{2\left( \left< R^>_a \right>\left< R^<_a \right> - \left< T^>_a
\right>\left< T^<_a \right>\right)}{ \left< R^>_a \right> + \left< R^<_a
\right> + \left< T^>_a \right> + \left< T^<_a \right>}
\right\}. \label{tok}
\end{eqnarray}
Here $\left< f \right> \equiv 2\int_{0}^{\infty} \left(-\frac{\partial
n_F(E)}{\partial E}\right) f(E) dE$, $n_F(E)$ is the Fermi distribution, and
the excitation  energy $E > 0$ is measured from the Fermi level;\\
$T_0^{>(<)}(E)$  is the transition probability
 for the  electron with energy $E$   incident from the left (right) reservoir
to  pass  to the right (left) one as an electron;  $T_a^{>(<)}(E)$  is  its
probability to reach the opposite reservoir as a hole (due to Andreev
scattering somewhere in between),
and $R_a^{>(<)}(E)$ is  the  probability  for  the electron incident from the
left (right) to be reflected into the same reservoir as a hole.

   In the weak coupling limit $\epsilon \rightarrow 0$
these transition and reflection  probabilities are given by the following
formulae;

\begin{eqnarray}
T_0^> \approx \frac{\epsilon^2}{|a_+|^2 |a_-|^2}\left( |a_+|^2 + |a_-|^2
+\right. \label{To>}
\\ \left.2{\tt Re} \left(\exp(i(\chi_h-\chi_e) a_- a_+^*
\gamma_{+,L}\gamma_{-,R}\right)\right);\nonumber\\
T_a^> \approx \frac{\epsilon^2}{|a_+|^2 |a_-|^2}\left( |a_+|^2 + |a_-|^2
+ \right.\label{Ta>}
\\ \left.2{\tt Re} \left(\exp(i( \chi_e- \chi_h) a_+ a_-^*
\gamma_{+,R}(\gamma_{+,L})^*\right)\right); \nonumber\\
T_a^< \approx \frac{\epsilon^2}{|a_+|^2 |a_-|^2}\left( |a_+|^2 + |a_-|^2
+ \right.\label{Ta<}
\\ \left.2{\tt Re} \left(\exp(i( \tilde{\chi}_h- \tilde{\chi}_e) a_+ a_-^*
\gamma_{+,R}(\gamma_{+,L})^*\right)\right); \nonumber\\
R_a^> \approx \frac{\epsilon^2}{|a_+|^2 |a_-|^2}\left( |a_+|^2 + |a_-|^2
+ \right.\label{Ra>}
\\ \left.2{\tt Re} \left(\exp(i( \chi_e + \tilde{\chi}_e - \chi_h -
\tilde{\chi}_h) a_+^* a_-
(\gamma_{+,R})^*\gamma_{-,R}(\gamma_{+,L})^2\right)\right); \nonumber\\
R_a^< \approx \frac{\epsilon^2}{|a_+|^2 |a_-|^2}\left( |a_+|^2 + |a_-|^2
+ \right.\label{Ra<}
\\ \left.2{\tt Re} \left(\exp(i( \chi_e + \tilde{\chi}_e - \chi_h -
\tilde{\chi}_h) a_+a_-^*
(\gamma_{+,L})^*\gamma_{-,L}(\gamma_{+,R})^2\right)\right). \nonumber\\
\end{eqnarray}
Here $\chi_{e,h} (\tilde{\chi}_{e,h})$ is the phase gained by an electron
(hole)  on the interval CD (resp. DC).

The  resonant denominators
\begin{eqnarray}
a_+ = 1 - t^4 \gamma_{+,L}\gamma_{-,R}\exp(i(\tilde{\chi}_e + \chi_h)),
\label{Alev1}\\
a_- = 1 - t^4 \gamma_{-,L}\gamma_{+,R}\exp(i(\chi_e + \tilde{\chi}_h))
\label{Alev2}
\end{eqnarray}
define the Andreev levels in the system.

As we see, the difference between various transmission and reflection
coefficients is only in the quickly oscillating terms (as function of electron
(hole) momentum). In the case of $N_{\perp}>>1$ channels their relative
contribution thus tends to zero, while it is quite insensitive to the
temperature averaging. They correspond in our case to the
 Spivak-Khmel'nitskii's osillations \cite{Spivak}.
Disregarding this difference we get from (\ref{tok}) a simplified expression:
\begin{eqnarray}
 G = \frac{2e^2}{h} \left\{ \left< T^>_0 \right> + \left< R^>_a \right>
\right\}. \label{tok2}
\end{eqnarray}

The behaviour of the transition and reflection probabilities
close to resonance is then governed by the "slow" terms:
\begin{eqnarray}
T_{0,a}^{>,<}(E) \approx R_a^{>,<}(E) \approx \epsilon^2
\left(\frac{1}{|a_+(E)|^2} + \frac{1}{|a_-(E)|^2} \right) \label{a+-} \\
\simeq \epsilon^2
\sum_{n,\pm} \left\{ \left(\frac{2L}{ \hbar
v_F}\right)^2 \cdot
\left( \left(E - E_n^{\pm}\right)^2 + \epsilon^2
\left(\frac{ \hbar  v_F}{L}\right)^2 \right) \right\}^{-1}. \nonumber
\end{eqnarray}
Here the sum is taken over the resonant values of energy $E_n$ are determined
from
(\ref{Alev1},\ref{Alev2}):
\begin{eqnarray}
E_n^{\pm} =  \frac{\pi \hbar  v_F n}{L} + \frac{ \hbar  v_F}{2L}
(\pi \mp  \Delta\phi).
\label{levels}
 \end{eqnarray}
This is the result obtained by Kulik \cite{K} for low-lying Andreev levels
in a long clean SNS junction,
but now due to leakage to the normal reservoirs, these levels acquire
finite wifth $\epsilon  \hbar  v_F/ {\cal L}$.

Substituting (\ref{Alev1},\ref{Alev2}) in (\ref{tok2}) and keeping only
the  terms  within the accuracy of $\epsilon^2$, we obtain a
 formula
\begin{eqnarray}
G =   \frac{2e^2}{h} \frac{ 2\epsilon^2}{\cos \phi_J +  1 + 2\epsilon^2}.
\label{G}
\end{eqnarray}
Here $\phi_J = \phi_B-\phi_A + 2\Phi/\Phi_0$ is the gauge invariant Josephson
phase difference between the ends of a link BDCA (see Fig.\ref{f.1}) in the
presence of magnetic flux $\Phi$ penetrating the loop BDCA; $\Phi_0 = hc/e$ is
the magnetic flux quantum. [We had to keep the term $2\epsilon^2$ in the
 denominator of (\ref{G}),
 since it is important close to resonance, when $\cos \phi_J +  1
\approx 0$.]

As we can see, the normal conductance through the system has strongly resonant
character (Fig.\ref{f.3}a). While we should expect the conductance of the order
of $(2e^2/h) \epsilon^2$ (since the probability for an electron/hole to enter
or leave the system at each of two junctions is of order $\epsilon$), the
actual quantity has sharp peaks, where the conductance reaches its maximum
possible value,    $2e^2/h$.

We can calculate the Josephson current in our system as well, e.g.,
using the Bardeen-Johnson approach \cite{BJ}, which allows us to express it
through the excitation spectrum at low energies. Specifically, the
Josephson current in a one-mode case can be written as follows:

\begin{eqnarray}
j_J = \frac{2e p_F}{\pi  \hbar } v_s  - \frac{2e v_F}{\pi {\cal L} p_F}
\sum_{E_n>0}
p(E_n) n_F(E_n - p(E_n)v_s).
\label{B}
\end{eqnarray}
Here $v_s =  \hbar  \phi_J / 2 m {\cal L}$  is the supercurrent velocity. The
first term in (\ref{B}) describes the (unperturbed) superflow due to
superconducting phase gradient $\phi_J / {\cal L}$, and the second term
is the contribution from the excited states (quasiparticle current).
The interplay of these contributions gives rise to the characteristic
"sawtooth"
shape of the $j_J(\phi_J)$ dependence in a long clean SNS junction \cite{BJ}.
It is important that at low temperatures only the lowest Andreev levels $E_n$
in (\ref{B}) are occupied and contribute to the current, and therefore we can
use in this expression the low-energy approximation for the levels $E_n$
(\ref{levels}), taking into account their finite width as well. The resulting
expression is (at $T=0$)
\begin{equation}
j_J^{(\epsilon)} = \frac{2e v_F}{\pi{\cal L}} \sum_{n=1}^{\infty} (-1)^{n+1}
e^{-2n\epsilon}
\frac{\sin
n\phi_J}{n} \label{jJ}
\end{equation}
(${\cal L} = L+2l$ is the total length of the wire BDCA). In the limit
$\epsilon \to 0$ we, of course, reproduce the  sawtooth dependence \cite{S,BJ}.

If compare the expression (\ref{jJ}) to (\ref{G}) (Fig.\ref{f.3}b),
we see that the   conductance at $\epsilon
\rightarrow 0$ actually behaves (within the accuracy of $\epsilon^2$) as
 the derivative of Josephson current with respect to the phase:

\begin{equation}
G =
\epsilon \left\{\frac{e^2}{\pi  \hbar }  -    \frac{e {\cal L} }{
 \hbar  v_{F}} \frac{d j_J^{(\epsilon)}}{d \phi_J}\right\}\label{Gphi}
\end{equation}
(cf. Eq.(\ref{main}) of this paper).
This result  stresses the fact that while the maximum conductance is achieved
when the Andreev levels in the system are tuned to the Fermi level
(\ref{To>}-\ref{Ra<}), the Josephson current is carried by the very Andreev
levels (see, e.g., \cite{Beenakker}).  The effects of this in the enhancement
of the normal conductance Aharonov-Bohm oscillations in mesoscopic rings  in
contact with superconductor will be discussed elsewhere.

The above  results are  directly generalized to the case of $N_{\perp} > 1$
non-mixing transverse modes.  Indeed,  the resonant denominators in
Eq.(\ref{G}) now  acquire dependence on the transverse energies, $E_{\perp,
\nu}$, and we have to sum  over the transverse mode indices, $\nu$ (or
integrate over $dE_{\perp}$, if $N_{\perp} \gg 1$). It is easy to see that in
either case the result will be simply $N_{\perp}$ times the 1D conductance, if
only $k_B T < \Delta E_{\perp}$. (In a realistic case of 1000 \AA   wide Ag
wire $\Delta E_{\perp} (\simeq 2 {\rm K}$.) On the other hand, the Josephson
current in a clean SNS contact is directly proportional to its area, i.e. to
$N_{\perp}$,
so that the relations (\ref{main1},\ref{main}) hold (we should only instead
of the longitudinal Fermi velocity $v_F$ write its average
over $N_{\perp}$ transverse modes, $\bar{v}_F = 1/N_{\perp}
\sum_{\nu=1}^{N_{\perp}} v_{F, \nu}$).

It is noteworthy that the amplitude of the phase-dependent conductance
oscillations can now significantly exceed $e^2/h$, which is its characteristic
value in dirty SNS systems \cite{Spivak,Alt}. This is the manifestation of the
ballistic motion  of the excitations in the system.

 Now we would like to suggest a direct experimental application of the result
Eqs.(\ref{main}\ref{Gphi}). Namely, it allows a direct
 measurement  of the phase dependence of Josephson current in a metallic wire
by measuring the equilibrium thermal noise in it (Fig.\ref{f.4}a). Indeed, by
virtue of Nyquist theorem \cite{Book} the spectral density of equilibrium
voltage noise is
\begin{equation} \langle \delta V^2 \rangle_{\omega} = 4 k_B T R,
\end{equation}
where $R = 1/G$ is the normal resistance of the system.
 The resonant dips in the noise level  will then follow the phase derivative of
the Josephson current through the normal bridge (Fig.\ref{f.4}b).

Let us  estimate the conditions under which such a measurement is possible.
The width of the Andreev level in the system, of order $\epsilon \hbar
v_F/{\cal
L}$,
 must be larger than temperature, i.e.,
\begin{equation}
\frac{T}{T_c} < \epsilon \frac{\xi_0}{\cal L},
\end{equation}
where $\xi_0 \sim  \hbar  v_F/ T_c$ is the superconducting coherence length.
This gives an estimate $T <  2$K for $\epsilon = 0.1, \;\; T_c = 10$K and
micrometer-sized system.
The thermal voltage  noise  intensity is then confined to an interval
\begin{equation}
{\epsilon} \cdot \frac{2 k_B T_c \xi_0 h}{ {\cal L} e^2} \leq  \langle \delta
V^2 \rangle_{\omega} \leq \frac{1}{\epsilon} \cdot \frac{2 k_B T_c \xi_0 h}{
{\cal L} e^2},
\end{equation}
which for $T_c \sim 10$K, ${\cal L} \sim \xi$  and $\epsilon = 0.1$ yields
\begin{equation}
7 \cdot 10^{-19} \frac{{\rm V}^2}{\rm Hz} \leq  \langle \delta V^2
\rangle_{\omega} \leq 7 \cdot 10^{-17} \frac{{\rm V}^2}{\rm Hz},
\end{equation}
which  can be measured experimentally.

In conclusion, we have investigated the normal current in a quasi1D mesoscopic
system, controlled by the phase difference in the superconductor, connected to
it. The magnitude of the conductance oscillations can significantly exceed the
conductance quantum, $2e^2/h$. We have shown also that the normal current
contains a term  which behaves as  the phase derivative of the Josephson
current in the system. This allows to measure directly the dependence of the
Josephson current on the superconducting phase difference and magnetic field,
by measuring the equilibrium voltage fluctuations in the normal part of the
system. The size of the effect is estimated.

\acknowledgements

The authors are grateful to V. Antonov, C.W.J. Beenakker, T. Claeson, P.
Delsing, Yu. Galperin, L. Gorelik,
V. Petrashov, S. Rashkeev, V. Shumejko, and A. Slutskin for many fruitful
discussions.

This work was supported by the Royal Swedish Academy of Sciences (KVA), the
Swedish Natural Science Research Council (NFR), and by  the Swedish Board
for Industrial and Technical Development (NUTEK). One of us (A.K.) gratefully
acknowledges the hospitality of the Department of Applied Physics and
Department of Physics, Chalmers University of Technology and G\"{o}teborg
University.

\begin{figure}
\caption{The schematic view of the ballistic normal metal system in contact
with the superconductor and thermal reservoirs (L and R). The scattering occurs
only
at the NS boundary (A and B) and in the nodes (C and D).}\label{f.1}
\end{figure}

\begin{figure}
\caption{The effective scattering matrices in the nodes C, D.}\label{f.2}
\end{figure}

\begin{figure}
\caption{Josephson current and normal conductance of the system as functions of
the superconducting phase difference between A and B, $\phi_J$.}\label{f.3}
\end{figure}

\begin{figure}
\caption{Suggested scheme of equilibrium noise measurement in  a clean normal
wire (a) and the noise dependence on $\phi_J$.}\label{f.4}
\end{figure}

\end{document}